# The Voltage Regulation of Boost Converters Using Dual Heuristic Programming

Sepehr Saadatmand
*Department of Electrical and Computer Engineering*
*Missouri University of S&T*
Rolla, Missouri, USA
sszgz@mst.edu

Mohammadamir Kavousi
*Department of Electrical and Computer Engineering*
*University of California, Riverside,*
Riverside, California, USA
mkavo003@ucr.edu

Sima Azizi
*Department of Electrical and Computer Engineering*
*Missouri University of S&T*
Rolla, Missouri, USA
sacc5@mst.edu

*Abstract*—In this paper, a dual heuristic programming controller is proposed to control a boost converter. Conventional controllers such as proportional-integral-derivative (PID) or proportional-integral (PI) are designed based on the linearized small-signal model near the operating point. Therefore, the performance of the controller during start-up, load change, or input voltage variation is not optimal since the system model changes by varying the operating point. The dual heuristic programming controller optimally controls the boost converter by following the approximate dynamic programming. The advantage of the DHP is that the neural network–based characteristic of the proposed controller enables boost converters to easily cope with large disturbances. A DHP with a well-trained critic and action networks can perform as an optimal controller for the boost converter. To compare the effectiveness of the traditional PI-based and the DHP boost converter, the simulation results are provided.

*Index Terms*— *Adaptive critic design, Boost converter, DC–DC converters, Model predictive controller, Dual heuristic programming, Reinforcement learning*

## I. INTRODUCTION

In the past few decades, power electronics DC–DC converters have matured into ubiquitous technologies. DC–DC power converters are used in a wide variety of applications, such as electronic devices like tablets and laptops, and in aerospace and power systems. The growth of renewable energy sources (RESs), such as uninterruptible power supplies (UPSs), wind turbines, and photovoltaics, has increased the interest on DC–DC power converters. The climate-based characteristics of the renewable energies sources lead to output voltage disturbances when facing load variations. Therefore, there has been a greater variety of research studies on the control scheme of DC–DC power converters. The three most important categories of DC–DC power converters include (i) buck, (ii) boost, and (iii) buck–boost [1]-[6].

To connect these energy resources to the grid, DC–AC inverters are used. However, the voltage level provided by several energy sources, such as photovoltaics and fuel cells, is lower than the required voltage for the inverter; therefore, the voltage level needs to be increased by boost converters. Boost converters, also known as step-up converters, are basic DC–DC converters that convert energy from the primary side to the secondary side by increasing the output voltage. An intermediate unit is used to connect residential photovoltaics into the grid. For these reasons, boost converters have attracted a large variety of attention [7], [8].

Controlling power electronics converters is a challenging concern because of their nonlinearity (hybrid) characteristics caused by the switching. In addition, specifically in boost converters with a right half-plane, stabilization is a concern. An undesired decrease in error bandwidth can overcome this drawback. Based on the control concept of boost converters, there are various categories as voltage control and current control, fixed frequency and unfixed frequency, linear or nonlinear controller [9].

The most common approach to controlling a boost converter is based on tuning the pulse width modulation (PWM) that controls the switch position. Conventional proportional-integral-derivative (PID) or proportional-integral (PI) controllers are the most common, thanks to their easy-to-implement characteristics. PI or PID-type controllers are designed based on the small-signal model of the averaging circuit. The small-signal model is the linearized model of the averaging circuit around a specific operating point. These types of controllers are designed for small perturbations, and their effectiveness is highly affected when facing a large signal disturbance. In other words, the performance of conventional-type controllers is not suitable when facing uncertainties or large disturbances [10].

The other popular controller for boost converters is known as sliding mode control (SMC), which was first introduced in [11]. The most highlighted feature of SMC is their inherent variable structure, and the most negative point is its variable switching frequency, which is a concern regarding electromagnetic interference (EMI) analysis [12]. Several studies have tackled the SMC approach to overcome its drawbacks and improve its performance. A PWM-based adaptive SMC is introduced in [13] that behaves like a traditional PWM controller with a fixed frequency; however, this method needs an auxiliary hysteresis block. An $H^\infty$ control is proposed in [14] to regulate a boost converter based on the sliding-mode current control. Although the SMC technique has several advantages, drawbacks such as EMI, chattering, and auxiliary blocks make it less compelling.

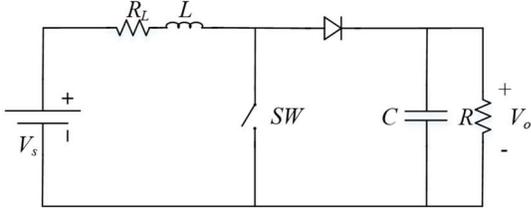

Figure 1. The circuit diagram of a boost converter

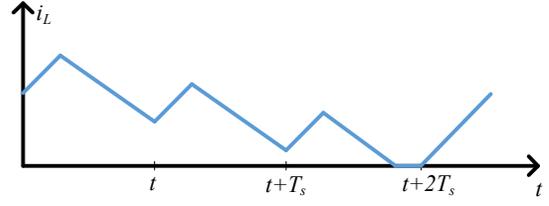

Figure 2. The inductor current mode describes the inverter mode: CCM mode when the inductor current is positive in $t \in [t, t + T_s]$, and it operates in DCM mode in $t \in [t + T_s, t + 2T_s]$

The enhancement in the state-of-the-art microcontroller and its affordability have increased the interest in nonlinear optimal controllers. Dynamic Programming (DP) and model predictive controller (MPC) have been implemented in different control applications. The first one derives an optimal law based on the Bellman's equation to optimize the cost-to-go function, and the latter minimizes the cumulative cost in a specific time horizon. Several studies have implemented MPC approaches [15], [16], but DP optimizer are hard to design and implement. Therefore, by the knowledge of the author, there have been no studies in implementation of DP DC–DC power converters. Approximate/adaptive dynamic Programming (ADP) tackles the drawback of DP by using artificial neural networks (ANNs) to solve the optimization problem. Adaptive critic designs (ACDs) are subcategories of ADPs. Dual Heuristic Programming (DHP) is a value gradient learning technique. ACD methods are used in power-frequency regulation of grid-connected virtual inertia-based inverters [17]-[19]. Besides the application of ACDs in DC/DC power converter are proposed in [20] and [21].

The main contribution of this paper is to propose a dual heuristic programming approach for the voltage regulation of a boost converter. The rest of the paper is organized as follows. Section II discusses the mathematical model of the boost converter. The dual heuristic programming, the training process, and implementation are explained in Section III. The simulation results are provided in Section IV to evaluate the performance and the effectiveness of the proposed controller. Lastly, the conclusion is presented in Section V.

## II. BOOST CONVERTERS

The circuit framework of a boost converter is shown in Figure 1. In this figure, *C, L,* and *R* are the output capacitor, input inductor, and the load resistor, respectively. Two power electronics switches are used: a controllable switch, *Sw*, and a diode, *D*. The output voltage, which is typically fixed, is shown by $v_o$, and the input voltage, which is typically variable, is shown by $v_s$. The internal resistor of the inductor is also shown by $R_L$. In this model, the diode on-time resistance, equivalent series resistance of the capacitor, and switch on-time resistance are ignored. The state-space model of the system in a continuous-time region is presented. The discontinuous-time state-space model can be easily derived from the continuous-time model. The small-signal averaging model is not discussed in this section because the proposed controller is designed based on nonlinear systems.

The independent state vector that represents the proposed boost converter includes two variables: (i) the inductor current and (ii) the output voltage (the voltage across the output capacitor) [22], which can be defined as

$$x(t) = [i_L(t) \ v_o(t)]^T. \quad (1)$$

Using the linear affine (linear plus offset), the proposed boost converter can be described by

$$\frac{dx(t)}{dt} = \begin{cases} A_1 x(t) + B v_s(t), & S = 1 \\ A_2 x(t) + B v_s(t), & S = 0, \text{ and } i_L(t) > 0 \\ A_3 x(t), & S = 0, \text{ and } i_L(t) = 0 \end{cases} \quad (2)$$

where the state matrices can be defined as

$$A_1 = \begin{bmatrix} -\frac{R_L}{L} & 0 \\ 0 & -\frac{1}{RC} \end{bmatrix} \quad A_2 = \begin{bmatrix} -\frac{R_L}{L} & -\frac{1}{L} \\ -\frac{1}{C} & -\frac{1}{RC} \end{bmatrix}$$

$$A_3 = \begin{bmatrix} -\frac{R_L}{L} & 0 \\ 0 & -\frac{1}{RC} \end{bmatrix} \quad B = \begin{bmatrix} \frac{1}{L} & 0 \end{bmatrix}^T. \quad (3)$$

There are two main categories regarding the operating point in boost converters: (i) continuous conduction mode (CCM) and (ii) discontinuous conduction mode (DCM). In CCM mode, the inductor current is always positive regardless of the switch position, but in DCM mode, the inductor current is zero for a period of time when the switch is off. Figure 2 depicts a situation when the boost converter can perform in both CCM and DCM.

## III. DUAL HEURISTIC PROGRAMMING

Neural networks have been used in various applications [23]-[26]. To optimize a control system over time neural network–based ACDs are suitable tools. The highlighted feature of ACDs is their ability to perform under conditions of noise and uncertainties. A category of adaptive critic designs by combining the reinforcement learning and dynamic programing is proposed in [27]. The action network and critic network are the main two important parts of a typical ACD. The action network and the critic network can be connected together through an identification model (model-dependent design) or directly (action-dependent design). Figure 3 depicts the block diagram of a simple ACD. The objective of the action network is to provide a series of control to optimize a utility function over time, and the critic network objective is to criticize how good the action network performs. There are four main classes for implementing ACDs known as dual heuristic dynamic programing (DHP), heuristic dynamic programing (HDP), global dual heuristic dynamic programing (GDHP), and global heuristic dynamic programing (GHDP). In this paper, a DHP-based controller is proposed and to perform the effectiveness a comparison with a PI controller is made.

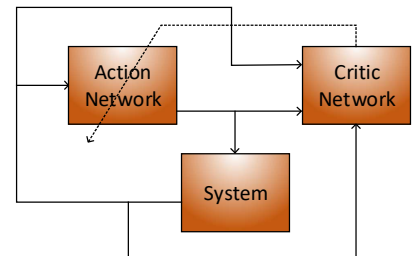

Figure 3. Adaptive critic design block diagram

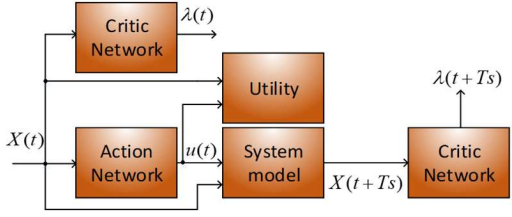

Figure 4. The DHP block diagram

Assuming that the optimal policy can be expressed as a differentiable function of the state variables, dynamic programing provides a set of control or a control policy to minimize the cost-to-go function defined as

$$J(t) = \sum_{k=0}^{\infty} \gamma^i U(t+k) \quad (4)$$

to guarantee that the cost-to-go function converges a discount factor ($\gamma$) is introduced ($0 < \gamma < 1$). The utility function is represented by $U(\cdot)$. By rewriting (6) in the form of Bellman's Recursion, it can be presented as follows:

$$J(k) = U(k) + \gamma J(k+1) \quad (5)$$

In ACDs, feeding the derivative of cost-to-go function with respect to the state variable is the main goal of the critic network.

In this paper the utility function is expressed based on the weighted error of the references, shown as

$$U(k) = \sqrt{K_P e_P^2 + K_Q e_Q^2 + K_f e_f^2} \quad (6)$$

where $e_f, e_Q$, and $e_P$, are the error signals for the frequency, reactive power, and the reactive power, respectively. The aforementioned error can be written as

$$\begin{aligned} e_P &= P_{set} - P \\ e_Q &= Q_{set} - Q \\ e_f &= f_g - f, \end{aligned} \quad (7)$$

and $K_f, K_Q, K_Q$ represent the frequency coefficient, the reactive power coefficient, and the active power coefficient, respectively. In other words, the proposed coefficients are the simple form of a weighted normalized expression that defines the importance of each signal. Mathematically solving the dynamic programing is complex and expensive. ACDs proposed a technique to provide the optimal control set to minimize $J(\cdot)$. In order to train the neural network, the derivative of the error or cost-to-go function is needed to criticize how well the action network is functioning. For example, the critic network in HDP method estimates the cost function and then by taking its derivative, the feedback signal to the action network is generated.

Figure 4 shows the block diagram of a critic-based DHP controller. In this figure, the control vector is represented by $(t)$, which is produced by the action network, and the state vector is represented by $X(t)$. By feeding the action network control signal to the system/plant the next state vector can be measured/computed as $X(t + Ts)$, where $Ts$ is the sampling time. The main goal of the critic network is to generate the mandatory feedback to the action network to make sure the control outputs satisfy the plant control objective. As discussed, the main goal of the designed controller is to minimize $J(t)$, and in order to do that the critic neural network provides the action network with the gradient of $J(t)$ with respect to the state vector, shown with letter $\lambda$. In DHPs, the critic network estimates the cost-to-go derivatives with respect to the states directly.

### A. Critic neural network

As mentioned, the critic network objective is to estimate the gradient of cost-to-go function with respect to the system states. By taking the derivative of (5) as

$$\frac{\partial}{\partial X_i(t)} J(t) = \frac{\partial}{\partial X_i(t)} (U(t) + \gamma J(t+1)). \quad (8)$$

Due to the training of the critic network, the error signal is expressed as

$$\|Er\| = \sum_t e_c^T(t) e_c(t), \quad (9)$$

needs to be minimized over time period t. In (9), $e_c$ at each period is written as

$$e_c(t) = \frac{\partial}{\partial X(t)} J(t) - \frac{\partial}{\partial X(t)} (U(t) + \gamma J(t+1)). \quad (10)$$

In addition, by applying the chain rule in DHP, (10) can be written as follows:

$$\frac{\partial J(t+1)}{\partial X_j(t)} = \sum_{i=1}^{n} \lambda_i(t+1) \frac{\partial X_i(t+1)}{\partial X_j(t)} + \sum_{k=1}^{m} \sum_{i=1}^{n} \lambda_i(t+1) \frac{\partial X_i(t+1)}{\partial u_k(t)} \frac{\partial u_k(t)}{\partial X_j(t)} \quad (11)$$

where n is the number of states, m is the number of controls, and $\lambda\_i(t+1) = \partial J(t+1)/(\partial X\_i(t+1))$. In this paper, state vector is defined as $X = [P \ Q \ e_p \ e_q \ e_f \ \theta_i]$ and the control signal is the inverter voltage magnitude. By implementing (11) in (10), it can be expressed as

$$e_{c_j}(t) = \frac{\partial J(t)}{\partial X_j(t)} - \frac{\partial}{\partial X(t)} (U(t) + \gamma J(t+1)). \quad (12)$$

Equation (12) is used to train the critic network. Due to the right-hand side evaluation in (12), the exact system model is needed to compute the partial derivative of next state with respect to the current state. To do so, there are two solutions. First, if the system model in known and all the parameters are certain, the derivative can be directly computed. Second, if the system parameters are not certain, a neural network can be used as a system identifier. By training the system neural network, the aforementioned derivative can be computed and used. In this paper, it is assumed that the parameters are not certain and a pretrained fully connected forward neural network with two hidden layers consisting of five nodes at each layer is used to model the system.

### B. Action neural network

Generating a series of control signal for the immediate future to minimize the cost-to-go function is the main objective of the action neural network. In this paper, the goal is to minimize the cot-to-go function for a time horizon of 1000 m.sec. The implementation of the proposed action network is analogous to the implementation of the critic network. To implement the action network a fully-connected multi-layer feedforward neural network is selected. This neural network includes two hidden layer and there are 8 neuron at each layer. State vector is feed as the input signal to this network. The voltage magnitude of the inverter is the output of the action network, which goes to the PWM unit. The backpropagation technique is utilized in order to update weights in the action NN. The goal is to optimize $J(k)$ as follows:

$$\zeta = \sum_k \frac{\partial J(k+1)}{\partial u(k)}. \tag{13}$$

The gradient of $J(\cdot)$ is given by the critic network. This gradient is used to updates the weights of action neural network.

## IV. SIMULATION RESULTS

Recently, the fuel cell generation structures have fascinated a great variety of considerations because of their exclusive advantages such as high efficiency, no moving part, environment friendly, greater durability, and sustainability. Varying output voltage during the load changes can cause complicated control problems. Therefore, a stable boost converter is essential that utilizes the fuel cell energy with higher efficiency and satisfies the conditions of a cascaded DC-AC converter applications. To evaluate the proposed controller, a DHP -based controller is implemented to regulate a boost converter. The block diagram of the proposed controller is depicted in Figure 6. As shown in this figure, both PI and DHP are implemented. The DHP signal is disabled when the critic neural network is pretrained. In other words, the state signal goes to the PI controller, and this controller adjusts the output voltage. After utilizing the boost converter with random references of output voltage and load current, the training data (including the state and the duty cycle at each time step) is generated. After the critic network is pretrained, the DHP -based controller goes online and controls the boost converter. The action and critic networks are updated at each control cycle. This control scheme includes both offline (to pretrain the critic network) and online learning (online training process of both critic and action networks).

Table I. Boost converter parameters and information

| Parameter | Symbol | Value |
| --- | --- | --- |
| Input voltage | $V_s$ | 60 ± 10% V |
| Output voltage | $V_o$ | 200 V |
| Output power | $P_{out}$ | 500 ±60% W |
| Load resistor | $R$ | 50 –200 Ω |
| The resistance of inductor | $R_L$ | 0.5 Ω |
| Switching frequency | $f_{sw}$ | 20 kHz |
| inductor | $L$ | 860 μH |
| capacitor | $C$ | 860 μF |

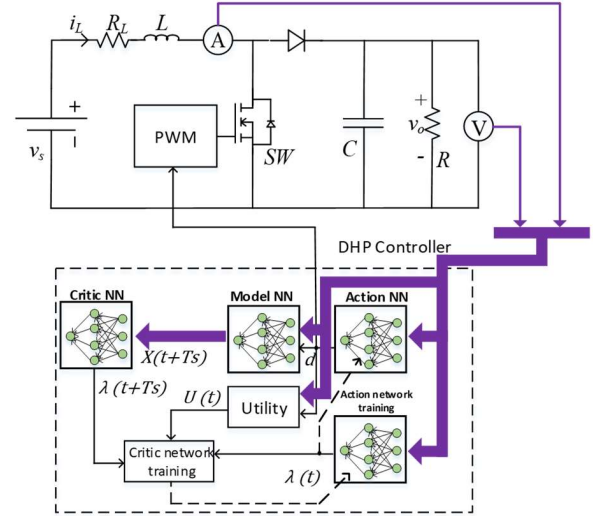

Figure 6. The block diagram of a DHP -based boost converter

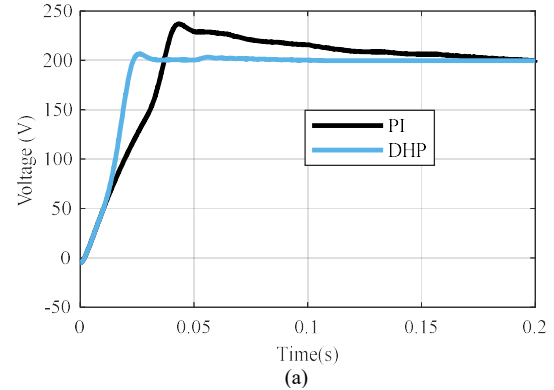

(a)

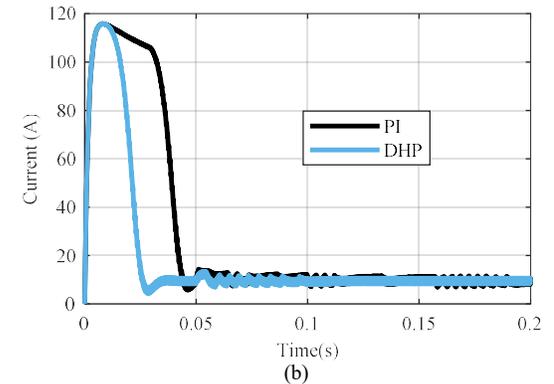

(b)

Figure 6. (a) The output voltage regarding the PI and DHP controller, (b) the inductor current regarding the PI and DHP controller

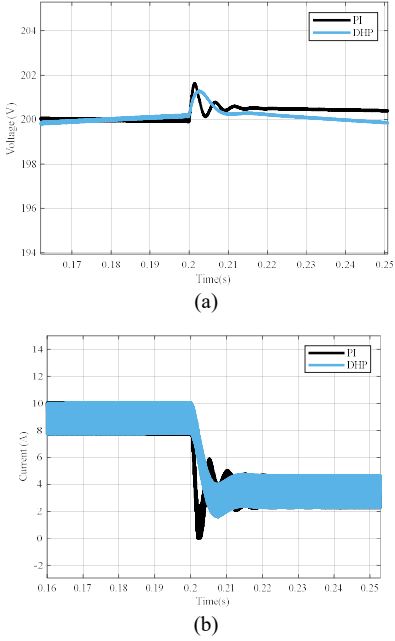

Figure 7. The performance of the boost converter in the change load resistance, (a) the voltage of the capacitor, (b) the current of the inductor

Table I illustrates the parameters of the proposed boost converter. The performance of the boost converter at start-up, load change, and the change of the input voltage is evaluated, and a comparison between DHP and a PI controller is shown.

### A. Performance In The Start-up

In this part the behavior and effectiveness of the proposed controller during the start-up is presented and is compared with a PI-based controller. The start-up is under the nominal load (i.e., $P_{out}$ = 500 W, $R$= 80 Ω). Figure 6 illustrates the output voltage and the current of the inductor of the proposed boost converter during start-up for the DHP and PI controller, respectively. As expected, the system does not operate in its nominal operating point during transient time. As shown, the DHP controller performs much quicker, and the settling time regarding the DHP is $t_{set}$ ≈ 5 msec, but the settling time regarding the PI controller is greater than $t_{set}$ ≈ 20 msec. The voltage overshoot regarding the DHP controller (3%) is much less than that of PI controller (18%).

### B. Load change

To assess the effectiveness of the proposed controller, a step-up load change scenario from 80 Ω to 200 Ω is simulated. As previous simulations show, the PI controller does not function well when the performance of the boost converter is not near the nominal operating point. Figure 7 illustrates the output voltage and the inductor current of the boost converter under both DHP controller and PI, respectively. As shown, the DHP controller keeps regulating the voltage optimally, but the stable PI controller starts oscillating after the change in operating point.

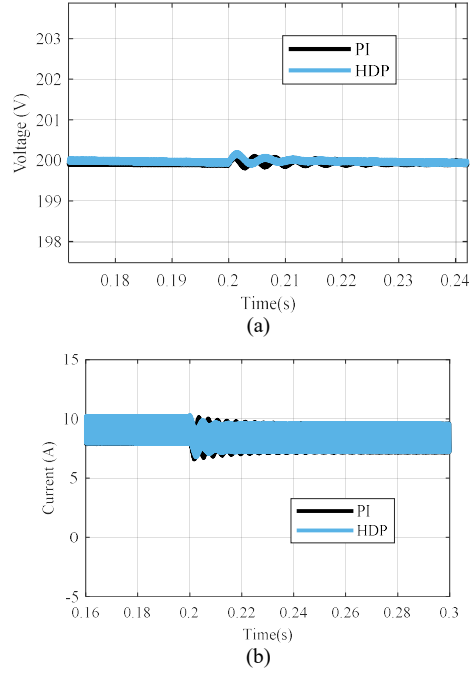

Figure 8. The performance of the boost converter in the reference voltage change, (a) the output voltage in reference voltage, (b) the inductor current

### C. Input voltage change

To evaluate the performance of the proposed controller regarding the input voltage changes, the maximum of reference voltage change is applied. The input voltage drops from 60 V to 54 V. Changing the reference voltage alters the linearized state-space model based on which the PI controller is designed. Therefore, the performance of the PI controller is not optimal. However, the DHP tracks the voltage reference with the minimum cumulative error at the optimal time horizon. Figure 8 depicts the voltage and the current output for both scenarios and for DHP and PI controller, respectively.

## V. CONCLUSION

In this paper, a heuristic dynamic programming (DHP) approach was introduced to control a boost converter optimally. The model-free and neural network–based characteristics of the DHP algorithm enable the controller to perform with more robustness when facing large disturbances. The drawbacks of the conventional PI/PID controllers have been discussed facing large disturbances. A well-trained DHP controller can regulate the output voltage of a boost converter. The performance of the proposed DHP-based and PI controller is compared via simulations. The DHP controller exhibits a voltage regulation with more robustness and faster dynamics compared to traditional PI-based boost converters. By validating the effectiveness of the proposed controller in three different scenarios (i.e., during start-up, load change, and input voltage variation), the proposed controller is introduced as a state-of-the-art control technique for boost converters.